\def\ff{$\varepsilon$}
\def\mion{$\rm M_i$}
\def\mc{M$_{\rm c}$}
\def\etal{{\it et~al. }}

\def\ms{$M_{\odot}$~}

\def\plane{${\rm log} T_{\rm eff} - {\rm log} L/L_{\odot}$ }

\documentstyle{laa}
\begin{document}

\title{Filling factors and ionized masses in planetary nebulae}

\author{Francesca R. Boffi\inst{1} \& Letizia Stanghellini\inst{2}}

\institute{Dept. of Physics and Astronomy, University of Oklahoma,
Norman, OK 73019, USA
\and
Osservatorio Astronomico di Bologna, via Zamboni 33,
                40126 Bologna, Italy}
\thesaurus{07.24.1 07.16.1  16.02.1 }
\offprints{Francesca R. Boffi}
\date{}
\maketitle
\begin{abstract}
We calculate filling factors (\ff) and ionized masses (\mion)
            for a total of 84 galactic and extragalactic planetary
            nebulae (PNe) at known distances.
	    To do these calculations,
            from the equation of energy propagation within a gaseous
            nebula, we have chosen forbidden line electron densities,
            observed angular diameters, and H${\beta}$ fluxes,
            from the most recent measurements available
            in the literature. Statistical analysis
            on the distributions of ${\varepsilon}$ and
            ${\rm M_i}$ show that (1) the ranges of values
            of these parameters is wider than what was previously
            found;
            (2) the mean value of the filling factor is
            between $0.3$ and $0.4$, for the different sets;
            (3) the mean value of the ionized mass is between
            $0.1$ and $0.25$ \ms, $0.2$ \ms representing
 an upper limit to the ionized mass for
		galactic disk PNe, and a typical value for galactic bulge and
extragalactic PNe;
            (4) a clear correlation between \ff~ and the dimensions of the PNe
was not found when distance-independent sets of PNe were used;
(5) for extragalactic PNe, where distance errors are not a factor, the
filling factors and the ionized masses anticorrelate tightly with
the electron densities. The results indicate that the modified Shklowsky
distance method is correct.

\keywords{Galaxy (the): the Bulge of -- Galaxies : Magellanic
               Clouds -- Planetary nebulae : general }

\end{abstract}

\section{Introduction}

Statistical studies of planetary nebulae (PNe) have been largely developed
in the recent years, giving results of great significance toward the
knowledge of the final stages of stellar evolution. In particular,
the analysis of samples of PNe at known distances, such as galactic bulge
and Magellanic Cloud PNe, have been used to derive an increasingly complete
picture of the overall nebular properties (e. g. Stasi\'nska \etal
1991, heretofore STAS91,
Dopita 1992). Nonetheless, in many of these statistical studies
the important aspect of the filling factor (${\varepsilon}$)
has often been overlooked, generally
because of the impossibility of estimating the exact value of this parameter
for each PN of a large set.

The filling factor
is a fundamental parameter, as it labels each individual PN
by telling which fraction of the nebular volume is filled by
ionized gas.
It has been introduced by Osterbrock \& Flather
(1959) to explain the existing discrepancies between observational
data and theoretical models in the case of the Orion nebula.
Usually, in statistical studies, the filling factor is arbitrarily assumed to
be
constant for all PNe (for galactic disk PNe, a value
equal to ${0.65}$ has been adopted (e.~g. Kaler 1970) and a value
equal to ${0.7}$ or $1$ is generally taken for the extragalactic
PNe (Wood \etal 1986; Dopita \etal 1988).
We do believe that this cannot be
a satisfactory approximation.
In fact, PNe do
come from progenitors having a mass comprised in a wide range
of values (${0.8}\leq{M_i}\leq8$ \ms, Iben \&
Renzini 1983) and are affected by
mass loss processes
occurring at different rates according to their mass
(Renzini 1989).
PNe show different morphological structures
and different degrees of ionization.
Only the direct calculation of ${\varepsilon}$ may provide a more
realistic study of PNe.
For the same reasons assuming a constant value for the ionized mass
of a set of PNe could be severely misleading.

Pioneer work on the subject of the filling factor has been performed by
Seaton (1966), who obtained the filling factor of fourteen
planetary nebulae from
photographic plates and drawings (average \ff~=0.63).
Later Webster (1969) found an average value of $0.8$ for
forty--nine PNe by using details of surface
brightness and PNe dimensions.
Similar values were obtained by O'Dell (1962). On the other hand,
Torres--Peimbert \& Peimbert (1977) calculated
much smaller values of ${\varepsilon}$.
More recently,
Mallik \& Peimbert (1988) have calculated the filling factor
of 35 galactic PNe at known distance (independent
from statistical methods) and found that ${\varepsilon}$,
calculated to be between $0.001$ and $1$,
anticorrelates with the nebular dimensions.
The average value of ${\varepsilon}$ found by Mallik \& Peimbert (1988)
was $0.28$.
Very recently Kingsburgh \& Barlow (1992) and Kingsburgh \&
English (1992) calculated filling factors for different types of
galactic nebulae, finding an average ${\varepsilon}$ of
of about 0.35 when excluding peculiar objects with ${\varepsilon}>1$.

On these grounds, we propose to extend the calculation of
${\varepsilon}$ and ${\rm M_i}$ to several sets of galactic and
extragalactic PNe whose distances are known independently
from statistical methods. The main goal of this paper
is to set constraints to ${\varepsilon}$ and ${\rm M_i}$,
to define the mean and most probable value for both parameters,
and to examine possible differences among the distributions of these
parameters in different stellar populations.
In \S 2 we set the theoretical formulation to calculate ${\varepsilon}$
and ${\rm M_i}$.
In \S 3 we present the input physical parameters of our four sets of PNe,
including galactic PNe, galactic bulge PNe,
Large Magellanic Cloud PNe and Small Magellanic Cloud PNe.
Results and correlations are reported in \S 4.
In \S 5 we draw conclusions and future perspective of this study.

\section{Theoretical formulation}

In order to set the stage for our calculations
let us consider a hydrogen--rich
nebula (we chose y=N(He)/N(H)=0.1) which gets ionized by a star.
The nebula is in thermal balance, and each ionization is balanced by
a recombination.
The equation that describes the propagation of energy
within this gaseous nebula can be written as:
$${\rm F_{ H\beta }} ={\left( {\rm {\alpha}_{\beta}}
{\rm h} {\rm {\nu}_{\beta}}\right) \over
\left(4 {\pi} {\rm d^2}\right)} \int_0^{\rm R_{N}}
{\varepsilon} \cdot {\rm N_{e}(r)}
{\rm N_{p}(r)}\cdot 4 {\pi} {\rm r^2} {\rm dr}~~~~
{\rm erg ~~cm^{-2} s^{-1} }, \eqno(1)$$
where
${\rm F_{ H\beta }}$ is the Balmer ${\lambda 4861}$ \AA~
flux corrected for extinction,
${\rm {\alpha}_{\beta}}$ is the hydrogen recombination coefficient,
${\rm \nu_{\beta}}$ is the frequency of the ${\rm H{\beta}}$ line,
d is the distance to the PN,
${\rm N_{e}(r)}$ is the electron density,
${\rm N_{p}(r)}$ the ion density,
and ${\varepsilon}$ the filling factor, that represents
which fraction of the nebular volume is occupied
by ionized gas.
The integration is performed over the whole volume of the nebula,
${\rm R_N}$ being the nebular radius.
The recombination coefficient can be written as:
$${\rm {\alpha}_{\beta}}={{9.69}\cdot{10}^{-11}}\cdot{\rm T^{-0.88}_{e}}
{}~~~~~{\rm cm^3 ~~s^{-1}}, \eqno(2)$$
which represents an interpolation from values found in the literature
(Brocklehurst 1971, Table V).

In the approximation that the density is constant
throughout the nebula,
from Eq. (1) we write:
$${\varepsilon}\cong{{2.47}\cdot{10}^6}\cdot
{{\rm F_{H\beta}}}{
\left(1\over {\rm {\alpha}_{\beta}}{\theta}^3
{\rm N^2_{e}} d \right)}. \eqno(3)$$
In Eq. (3) the approximation of fine angles holds
(i.e. the nebular radius ${\rm R_{N}}\approx{\rm {\theta}} {\rm d}$,
where the angular radius $\theta$ is expressed in ${\rm arcsec}$ and
the distance in ${\rm kpc}$) and it has been assumed that
$$k={{\rm N_{e}}\over{\rm N_{p}}}\simeq{1.15}, \eqno(4)$$
as to keep into account partial double--ionization of helium atoms.

To calculate the ionized masses we assume simple spherical
geometry,
$${\rm M_{i}}={{4 \pi {\rm R^3_{N}}\over3}}\cdot{{\varepsilon}
{\rm N_{e}}  {\rm m_{p}} {{\rm 1+4y}\over{\rm k}}}
{}~~~~~~{\rm g}, \eqno(5)$$
where y=0.1 and k=1.15.
If we use Eq. (3) for ${\varepsilon}$,
${\rm M_i}$ turns out to be independent on ${\theta}$:
$${\rm M_{i}}\cong {{3.45}\cdot{10^{-2}}} \cdot
\left({ {\rm F_{H\beta}} d^2} \over { {\rm {\alpha}_{\beta}}{\rm N_e}} \right)
{}~~~~~~~{\rm M_{\odot}}. \eqno(6)$$
We then use Eqs. (3) and (6) through our calculation.

\section{Input Database}

We calculate ${\varepsilon}$ and ${\rm M_i}$ for 84 PNe:
twenty-nine belonging to the Galaxy,
twelve to the galactic bulge,
nineteen to the Small Magellanic Cloud, and twenty-four to
the Large Magellanic Cloud.
Our PNe database includes only known distance objects, that have been selected
from the literature among those with electron density calculated by
forbidden line analysis, and whose ${\rm H\beta}$ fluxes and angular radii
are reliably known. Following, we examine the parameter space for each
set of PNe.

\subsection{Galactic Planetary Nebulae}

In Table 1 we list our choice of galactic PNe.
Column (1) gives the PN name.
In columns (2) and (3) the ${\rm H{\beta}}$ flux
(in units of ${\rm erg~~cm^{-2} s^{-1}}$) and
the extinction logarithmic factor,
for which the flux needs to be corrected, are reported.
Fluxes are already in the modern photometric scale as suggested
in Shaw \& Kaler (1982) and as explained in Cahn \etal (1992,
hereafter CKS92).

The electron densities quoted in column (4) of Tab. 1 are, for the most part,
obtained by averaging the density values
obtained from the different ions, as calculated by
Stanghellini \& Kaler (1989, hereafter SK89). When these densities
were not available, we have used other sources, as given in the
footnotes to the Table.
Densities from SK89 are preferable since they have been
calculated by using the most recent atomic parameters.
All SK89 densities listed in column (4) belong to the low error
domain of the intensity ratios--densities curves.
There are two galactic PNe for which the calculation of
${\varepsilon}$ and ${\rm M_i}$ has been performed, but their densities are
within the higher error domain, thus they have not been reported in the
table nor counted for the statistics.

In column (5) we give the electron temperature from Kaler (1986), or from
other sources listed in the footnotes.
We calculated a weighted average of the [N II] and [O III] electron
temperatures according to the electron densities that have been
used to get the mean electron density in column (4),
and by taking into account the fact that
temperatures derived from the intensity of [N II] lines
are well representative of regions of low ionization where
oxygen and sulfur are easily ionized,
and that those derived from the intensity of [O III]
are instead well representative of regions of high ionization,
where argon and chlorine are preferentially ionized
(Torres--Peimbert \& Peimbert 1977).
In the case that
no temperatures have been found in the literature,
we adopt a standard value of 10000 ${\rm K}$.
This value is very close to the real value of
these low density environments,
and represents the typical average value guaranteed
in the actual case of thermal balance (Osterbrock 1989).

In column (6) we give angular radii, and in column (7) the distances to the
nebulae which are for the most part reddening distances, the exceptions
being some non--LTE model-dependent distances and two cluster memberships
(see the footnotes to the Table; reference to CKS92 are relative to the
distances there reported on Table 3). Heretofore, we refer to Set 1
as the group of galactic PNe with model-independent distances, and to
Set 2 to these galactic PNe whose distances have been calculated by
Mendez \etal (1988, heretofore MEA88).
In columns (8), (9) and (10) we list the linear radii,
and the calculated filling factors and ionized masses.

\subsection{Galactic Bulge Planetary Nebulae}

We have chosen radio selected objects from the catalogs
by Gathier \etal (1983), and Pottasch \& Acker (1989), since the
radio observation helps to reduce the severe
extinction effects present in the case of objects
in (or in the direction of) the galactic
bulge (Habing \etal 1989).
We have taken all the PNe from these two catalogs
with no repetition.
There is good confidence that
the objects of the two catalogs are true PNe
(Pottasch 1983;
Gathier \etal 1983;  Pottasch \& Acker 1989).

Table 2 is the analogous to Table 1 for galactic bulge PNe.
The electron densities listed
in column (4) are
the means computed in the same way than for galactic PNe.
Electron temperatures, given in column (5),
are from Acker \etal (1989b).
In column (6) we list the angular radii,
from CKS92 or
Acker \etal (1989a).
The physical radii, filling factors and
ionized masses, respectively listed in the last three columns,
have been calculated with a distance to the galactic center of
${8.5}$ kpc (Reid 1989).

\subsection{Magellanic Cloud Planetary Nebulae}

There are two reasons why we selected Magellanic Clouds PNe.
First planetary nebulae of these
two external galaxies have been extensively studied in the last ten years,
and a wealth of information is ready for us to use.
Furthermore, as Dopita \& Meatheringham (1990) underline,
these objects are at known distance and
evolve in a low reddening environment.
Our database consist of 19 SMC and 24 LMC PNe, collected
from the catalog by
Sanduleak \etal (1978) (hereafter SMP78). In fact, these were the only
Magellanic Cloud PNe whose necessary physical parameters for the calculation
of $\varepsilon$ and ${\rm M_i}$ were available in the literature.

Tables 3 and 4, structured in the same way as Tab 2,
give the required input data and the calculated parameters
for Small and Large Magellanic Cloud PNe.
Columns (2) and (3) of Tabs. 3 and 4
give ${\rm H{\beta}}$ fluxes and logarithmic extinction
correction factors;
each flux value is an average value of all the values
found in literature from
Meatheringham \etal 1988,
Wood \etal 1987 (hereafter WMDM), Barlow 1987, Webster 1969 \&1983,
Aller 1983, and  Osmer 1976.
The WMDM fluxes have been rejected in the case of
SMP 2, 11 and 17 in SMC and of SMP 40 in LMC
since they are not in good agreement with the corresponding
Meatheringham \etal (1988) quantities (see for
further comments the latter paper).
All the foretold fluxes are not corrected for the
interstellar extinction and still need to be brought into
the modern photometric scale, by subtracting
$0.02$ (Shaw \& Kaler 1982).

The logarithmic extinction are
from Kaler \& Jacoby (1990) and Meatheringham \& Dopita
(1991).
Thanks to the fact that in the Magellanic Clouds
the extinction is very low, we have
assumed a constant logarithmic extinction factor
for those PNe whose extinction have not been measured.
We then adopt respectively c=0.12 for SMC PNe and c=0.21 for LMC PNe,
as explained
in detail in Kaler \& Jacoby (1990).
We test the validity of the assumed average values of
the extinction factor by estimating c through its definition, though using
the color excess of each Cloud calculated in Jacoby \etal (1990).
{}From the definition of c,
$${\rm c}={log\left( {\rm F_{H\beta}^{theor.}}\over{\rm F_{H\beta}^{obs.}}
\right)}={{\rm {A_{\lambda}} {\rm E(B-V)}}\over{2.5}}, \eqno (7)$$
where ${\rm F_{H\beta}^{theor.}}/{\rm F_{H\beta}^{obs.}}$
is the ratio of the corrected and observed ${\rm {H\beta}}$ fluxes,
E(B-V) is the color excess, and
${\rm A_{\lambda}}$ is a function of the wavelength.
In the case of SMC we obtain
c=0.09,
in good agreement with the adopted value,
and for LMC c=0.14.

The electron densities are
averages over all possible values reported in the
literature. For LMC and SMC PNe, the plasma diagnostics
has been performed only with the [O II] ${\lambda}{\lambda} {{3726}-{3729}}$
and the [S II] ${\lambda}{\lambda} {{6717}-{6731}}$ intensity ratios,
therefore the mean densities in columns (4)
of tabs. 3 and 4 are calculated
from the values from these ions, only as listed in
Dopita \etal (1988), Monk \etal (1988),
Meatheringham \& Dopita (1991), and Barlow (1987).
The average values of ${\rm N_e}$ from
Dopita \etal (1988) exclude those values
from the highest velocity component as to avoid peculiarities.

The electron temperatures are means of all possible values found
in the literature
(Monk \etal 1988, and Meatheringham \&
Dopita 1991). We have
taken temperatures calculated from the [N II] intensity
ratios since all electron densities come from the [O II] and
[S II] lines;
missing these temperatures, we have taken the ones
determined from the [O III] intensity ratios. A temperature of
10000 ${\rm K}$ has been assumed when no other reference was found.

In columns (6) of tabs 3 and 4
we list the angular radii. Only a few Magellanic Cloud PNe
have well measured radii, given the difficulty to achieve a good resolution of
these PNe from ground observations.
We have taken all angular
diameters of bright and compact objects ($r\leq{0.13}$ ${\rm pc}$)
from Wood \etal (1986) speckle interferometry data, while
those of faint and large PNe
(for angular diameters larger than $0.7$ ${\rm arcsec}$)
from WMDM 's direct imaging.
Some diameters come from Jacoby \etal (1990) and have been
derived with the method used in WMDM.
The mean angular diameters of
SMP 47, 62 and 78 in LMC are uncertain
as they have been calculated from limits with discordant signs.

Physical radii, filling factors and ionized masses are listed
in the last three columns.
The SMC distance modulus
is m-M=$18.8$, equivalent to a
distance of $57.5$ ${\rm kpc}$, while for LMC
we have chosen m-M=$18.5$, corresponding
to a distance of $50.1$ ${\rm kpc}$ (Kaler \& Jacoby 1990).
\begin{figure}
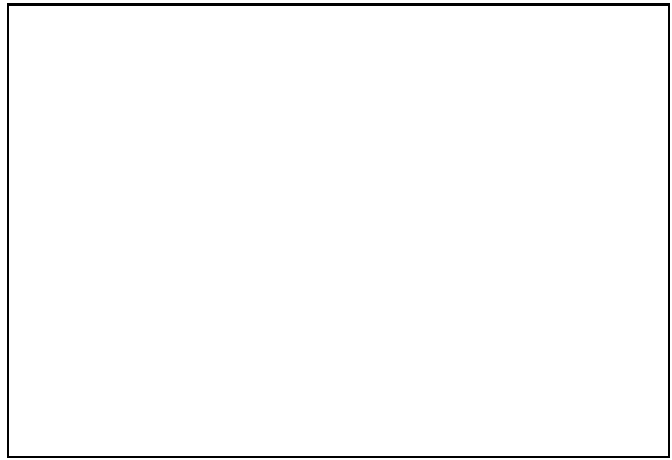

\picplace{6cm}
\caption[]{Cumulative histogram of the filling factors for the
four samples of Planetary Nebulae. Filled circles refer to
galactic PNe (Set 1); open circles to galactic bulge PNe;
filled squares to the LMC PNe; open squares to the SMC PNe.}
\end{figure}

\begin{figure}
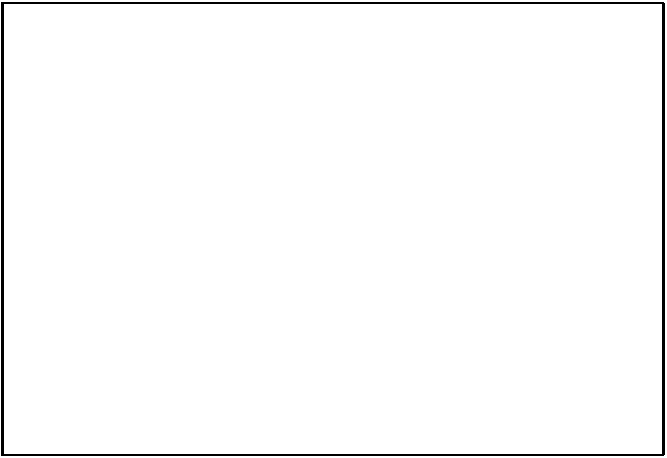

\picplace{6cm}
\caption[]{
Cumulative histogram of the ionized masses (in solar units) for the
four samples of PNe. The symbols used have the same
meaning as in Fig. 1.}
\end{figure}

\section{Results}

\subsection{Statistical analysis of the results}

In Figures 1 and 2 we show cumulative distributions of
${\varepsilon}$ and ${\rm M_i}$
for our four sets of PNe, where we restrict to values of the filling factor
between 0 and 1. In this chapter we always refer to
averages and correlations by excluding those objects
with ${\varepsilon}>1$. We see the off--limit cases in the next section.
Let us analyse in detail one set of PNe at the time starting with
galactic PNe.

Of the fourteen galactic PNe of Set 1,
eleven have ${\varepsilon}\leq1$ and for them
the mean ${\varepsilon}$
is equal to $0.37$
and the mean ${\rm M_i}$ is $0.11$ \ms
(averages of \ff~ and \mion~ for all sets of PNe are on Table 5).
The derived values of \ff~ for galactic PNe
are much smaller
than the ones usually adopted (see \S 1),
but very close to the ones calculated by Mallik \& Peimbert (1988),
and by Kingsburgh and collaborators.
{}From the cumulative histogram of Fig. 1 we see that
up to a value of about $0.35$ the distribution
of ${\varepsilon}$ in galactic PNe is systematically
lower than those of galactic bulge and extragalactic PNe.
This effect could be explained since late, low filling
factor nebulae are easily seen at larger distance, due to
prospective effects.

In Figure 2 we note a wide spread of the ionized masses, and that all bin
intervals are well represented.
A large percentage ($\sim 90\%$) of galactic PNe have ionized mass
between 0 and 0.2 \ms,
which means that the value of 0.2 \ms well represents
the upper limit to the mass of a typical disk
PN, while does not mean that all PNe can be well modelled by this
value of the mass.

In Figures 3 and 4 we show how ${\varepsilon}$ and
${\rm M_i}$ correlate
to nebular dimensions in galactic PNe.
In these Figures, circles refer to Set 1 and
squares to Set 2 (see \S 3). Some nebulae have been represented by two
symbols connected with a vertical line.
In these cases, the filled symbols always
refer to have taken into account electron densities within
the higher reliability
interval, while open symbols include in the means the offset values
(see SK89).
The correlations found and drawn in the Figures only refer to
the filled circles.
In both graphs two objects (NGC 5315 and He 2-131)
are plotted with an open circle only:
for them the only electron densities available in the literature
fall outside the higher reliability interval.
One object, NGC 4361, of Set 2, is not
plotted in the two graphs for scaling reasons
and it considerably
scatters from the general trend.

\begin{figure}
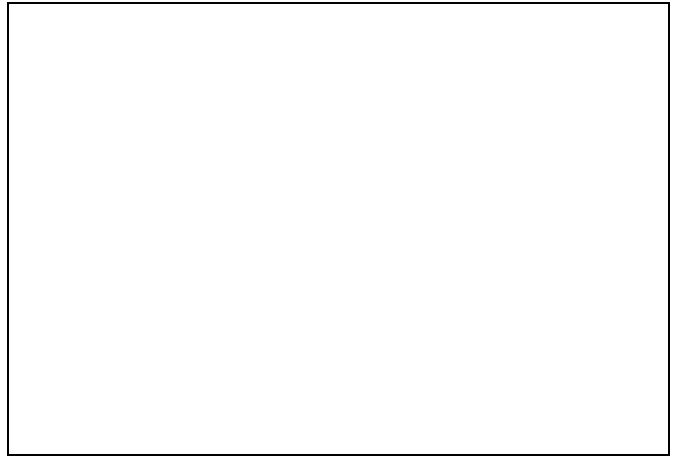

\picplace{6cm}
\caption[]{
Filling factor versus linear dimension in
galactic PNe:
filled circles=Set 1 (see text) taken with $\rm N_e$
within the range of higher reliability; open circles=Set 1 PNe
averaging all available electron densities;
similarly for filled and open squares, that refer to Set 2 PNe.
Vertical lines connect values for the same nebula when found for the different
assumptions on the electron densities. The solid line represent the
correlation between filling factor and dimension in Galactic PNe (see text).}
\end{figure}

By analyzing Fig. 3, we found
(for the PNe in Set 1, excluding those with ${\varepsilon}$ greater than 1)
$${{\rm log} \varepsilon=-1.29
-0.69 {\rm log R}},   \eqno (8)$$ with
correlation coefficient r=-0.4.
Our correlation
is weaker than what found by Mallik \& Peimbert (1988).
We feel that the use of more recent data have just disclosed that
the relation is not real, but it is due to data scatter,
in agreement with what found by
Kingsburgh \& Barlow (1992). Further support to this conclusion is given
by the analysis of galactic bulge PNe, later on this paper.

We then go on and check if there exists any
relation between the thickness of the nebulae and the
filling factor. We find that most optically thin planetary
nebulae (i.e. those with an optical thickness parameter $\rm T>{3.13}$, CKS92)
but one (NGC 6565) fall in the same region of the ${{\rm log}
\varepsilon}$ -- ${{\rm log} R}$ plane. In particular
we found that \ff~ $>$ 0.1 for optically thick nebulae.

\begin{figure}
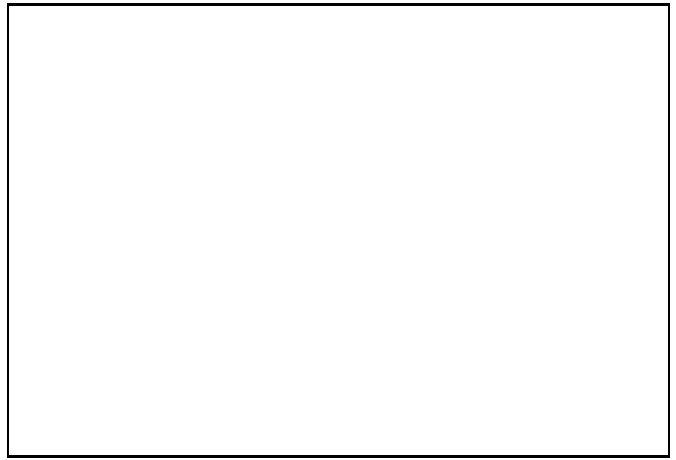

\picplace{6cm}
\caption[]{
log ${\rm M_i}$ vs. log R in galactic PNe;
The symbols have the same meaning
as in Fig. 3. The correlation takes into account filled circles only.}
\end{figure}

Figure 4 shows that ionized masses and nebular dimensions do
correlate, as $${\rm log} {\rm M_i}={0.188}+ {1.329}{\rm log R}, \eqno (9)$$
with regression coefficient r=0.72 ( the calculation was performed
for Set 1 PNe with ${\varepsilon}\leq1$ only).
All objects provided with model dependent distances (filled and open squares)
have systematically higher values of ${\rm M_i}$, as
they preferentially fall in the upper part of the graph where
${\rm M_i}\ge{0.15}$ \ms.
This result adds suspicion on the MEA88 distances,
and explains why we do not include these data on averages and correlations.

By comparing the results shown in Fig. 4 with a similar,
theoretical plot by
Schmidt--Voigt \& K\"oppen (1987b, Fig. 8)
we could think that our galactic PNe trace the 2W model the best.
Nevertheless, by looking at the different mass loss rates on the AGB
assumed by Schmidt--Voigt \& K\"oppen (1987a\&b), we conclude
that the observed data do not follow either relation (2W nor
3W) in such a fashion to allow us to decide for one or the other.

We now pass on galactic bulge PNe.
For these PNe, the average filling factor is equal to $0.39$,
very close to what found for galactic PNe, and the mean value of
${\rm M_i}$=0.22 \ms
is slightly larger than that of galactic PNe. From stellar evolutionary models,
we would expect that galactic bulge PNe, supposedly evolving
from Population II progenitors,
would have ionized masses
quite smaller than the masses of disk PNe (Iben \& Renzini 1983).
The high values of ionized masses that we found
could be due to different reasons: first, we can think to some selection effect
that allows a more probable detection of brighter PNe, which means
on average more massive PNe in direction of the galactic bulge (see also
Pottasch 1983);  second, massive objects, descended from progenitors
of a younger population than expected, do exist (Rich 1991); or, the
ionized masses of the galactic bulge PNe have been
overestimated in the calculation.

In particular the distance may have played a major role in this
sense, since ${\rm M_i}$ goes as the distance squared, and an overestimate in
the galactic bulge distance
by 20\% propagates into lowering the mean \mion~ of about 0.1 \ms.
Another possibility could be that the difference is produced
by the error propagation from all observed parameters.
In \S $4.2$ we see that this uncertainty can produce
a variation in the mass of $\sim{0.1}$ \ms, which could
explain the general trend in the differences.

\begin{figure}
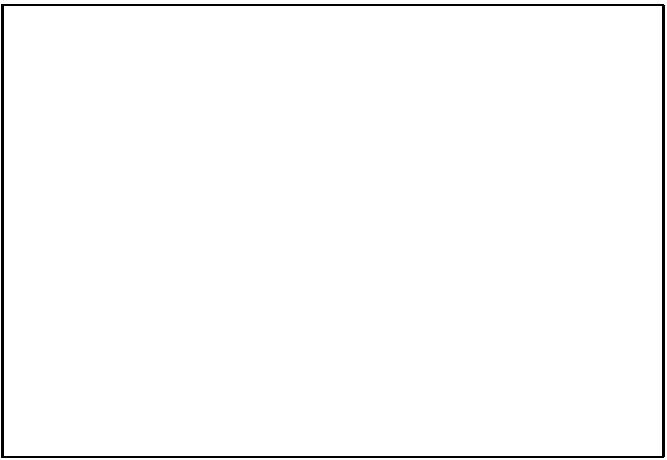

\picplace{6cm}
\caption[]{
log \ff~ vs. log R for galactic bulge PNe.}
\end{figure}

In Figure 5 we show that the correlation between log \ff~ and log R
is almost lost when using galactic bulge PNe.
We conclude that,
the data available up-to-date do not indicate correlation between the filling
factor and the nebular dimensions.
Further analysis with distance independent PNe samples is in order. In this
paper,
we could not go further in this direction, since the determination of angular
diameters for PNe of the Magellanic Clouds is still very uncertain; we propose
to investigate this important aspect in the future, with the new available data
on Magellanic Cloud PNe expecially from space observation.

At last, we go on to Magellanic Cloud PNe.
The Magellanic Clouds
mainly contain massive stars and young star clusters.
PNe not only may have descended
from an old star progeny, but also from massive progenitors
that have undergone a huge mass loss in AGB as to
be left with a low remnant mass well below the Chandrasekhar limit.
{}From Figure 2 we note that the mass spread in these galaxies is
even larger than for galactic PNe.
The masses of SMC PNe are in a range
${0}\div{0.8}$ \ms, with $47\%$ of the objects with ${\rm M_i} \le {0.2}$ \ms,
and the average mass value is ${0.24}$ \ms. Of the 19 objects of the SMC,
15 have filling factor smaller than 1,
the mean value of these being ${0.29}$.
On the other hand, LMC PNe show an average mass value
of ${0.21}$ \ms and a mean ${\varepsilon}$ of ${0.32}$.
Again the spread in mass values is large, going from
${0}$ to ${0.96}$ ${\rm M_{\odot}}$.

\begin{figure}
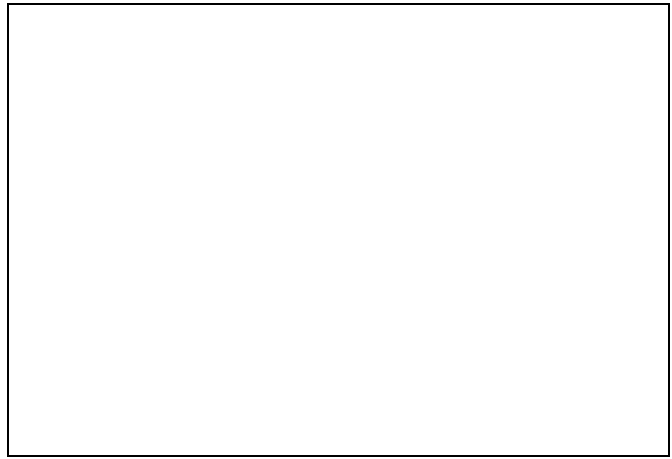

\picplace{6cm}
\caption[]{
log \ff~ vs. log ${\rm N_e}$ for Small Magellanic Cloud PNe;
The solid line represents the correlation found (see text).}
\end{figure}

\begin{figure}
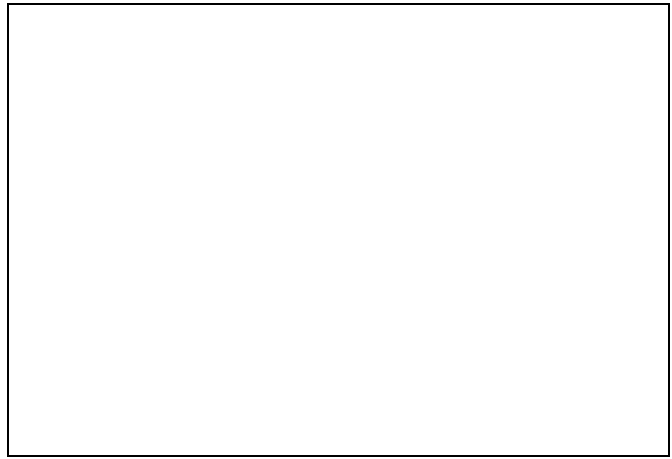

\picplace{6cm}
\caption[]{
log ${\rm M_i}$ vs. log ${\rm N_e}$ in Small Magellanic
Cloud PNe; the solid line represents the correlation found (see text).}
\end{figure}

Figure 6 shows that the trend between filling factor and electron density
in the SMC PNe: $${\rm log \varepsilon =3.41-1.14 log N_e},
\eqno (10) $$ with
r=-0.66, is weaker than expected from functional relation (Eq. 3).
This trend, that was only hinted in galactic PNe,
suggests that, as the expansion of the nebula proceeds,
the nebular material gets actually diluted.
In fact as the nebula expands
the volume increases while the electron density decreases.
The ionized material per unit volume
decreases being partially dispersed in the interstellar medium.

In Figure 7 we plot the relation between \mion~ and N$_{\rm e}$ for SMC PNe.
The logarithms of the two quantities seem to correlate well, and we found
$$ {\rm log M_i} = 1.88 - 0.70 {\rm log N_e}  \eqno (11)$$
with correlation  coefficient r=-0.78.
If we suppose that most SMC PNe are optically thick, we can test on this
Figure the validity  of Shklowsky's assumption
that the luminosity in the light of H$\beta$ is constant for all PNe.
If that were the case, \mion~ $\propto {\rm N_e}^{-1}$. Our relation is weaker
than this, as \mion~ $\propto {\rm N_e}^{-0.7}$. On the other hand,
for our data \mion~ is constant only for very low electron densities.
If we compare the last two Equations,
we can infer that \mion~ $\propto \varepsilon ^
{0.6}$, which is very close to the assumption under Shklowsky's modified
method (\mion~ $\propto \varepsilon ^{0.5}$), as used in CKS92, giving more
strength to the CKS92 distance scale.

We now examine the possible correlation between the ionized masses and the
masses of the parent stars.
for galactic PNe of Set 1 we use the central star
parameters as given in Stanghellini \etal (1993) and we locate the
central stars on the
\plane plane to infer their masses.
Given the large uncertainties on the distances and temperatures of these stars,
we could
not single out any correlation between the two quantities.

We then go on and test the same correlation in Magellanic Cloud PNe.
In Figure 8 we show ionized versus core masses (in solar units) for
SMC and LMC PNe. Open symbols refer to those nebulae
whose filling factor is larger than unity, those whose physical parameters
have the larger errors (see Sect. 4.2).
The ionized masses have been taken from Tabs. 3 and 4,
and the core masses from Kaler \& Jacoby (1990, 1991). Although the scatter
in Figure 8 is large, we can individuate two major
trends, expecially when we consider only the filled symbols: one group
of PNe show a large spread in \mion~ on a relatively narrow \mc~ range,
the other group has smaller \mion~, on average, on a wider \mc~ domain.
Before trying to interpret these results,
let us analyze the uncertainties in the
parameters.

\begin{figure}
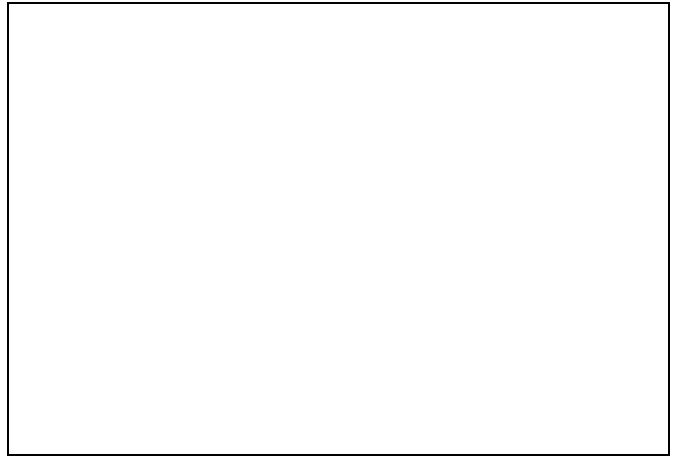

\picplace{6cm}
\caption[]{Ionized mass versus core mass (in solar units) for Magellanic Cloud
PNe.
Squares=LMC; Circles=SMC; open symbols: \ff $>$1; the lines correspond to the
PN mass-core mass relations by Vassiliadis \& Wood (1992), and Renzini \&
Voli (1981) cases A and B}
\end{figure}

The ionized mass approximates
reasonably well the total nebular mass for optically thin
nebulae, while M$_{\rm c}$
is an overestimate of the true value (see Kaler \& Jacoby 1990, 1991).
For optically thick PNe, on the other hand, \mion~ represents a lower limit to
the PN mass, while M$_{\rm c}$ is well determined.
This means that the points in Figure 8 could be mistaken as they have to
be shifted upward or leftward, depending on nebular thickness.
{}From Kaler \& Jacoby's (1990) criterion for optical thickness
(F($\lambda$3797
)/F(H$\beta$)$>$0.8 and 0.35 for LMC and SMC PNe respectively),
and from the flux data listed in the latter paper,
we found that most of the nebulae
in our sample are thick, thus the core masses in Figure 8 are sound values.
We chose three PNe for which the filling factor is less than unity, and
for which the thickness criterion by Kaler \& Jacoby (1990) is largely achieved
to indicate the typical errorbars (the PNe with errorbars
are SMP 21 and 40 in LMC and SMP 5 in SMC, for discussion on \mc~ errors see
Kaler \& Jacoby 1990).

{}From our error analysis we conclude that the two different trends
are real, thus a correlation between the core and the
planetary nebula masses can not be the same for all
types of PNe.
For more insight, we plot the relation between core and nebular
mass from Renzini \& Voli (1981, RV81); furthermore, from the calculation of
Vassiliadis \& Wood (1992, VW92), we derive another \mion~--\mc~ relation
by considering the amount of mass that has been ejected during the last
thermal pulse (or group of pulses in the cases where the mass loss is there
continuous, meaning that in case of a multiple shell PNe we only consider to
be observing the innermost shell).
Renzini \& Voli (1981) use high mass loss during AGB evolution
and earlier, while Vassiliadis \& Wood produce their models by using  mass
loss rates derived from empirical period--mass loss rate
relations of Miras and OH/IR stars.
We conclude that the different theoretical predictions encompass
all observed PNe, while no systematic discrepancy was found between the PNe
that
occupy different parts of the \mion~--\mc~ plane.

\subsection{Detailed analysis and peculiar objects}

In several cases the filling factor is calculated to be larger than unity.
Among galactic PNe, NGC 6565, NGC 6567, and NGC 6741 of Set 1, plus
NGC 6629 and IC 418 of Set 2 have \ff~ $>$1.
Except for the last nebula, where \ff~=0.89 if one takes into account electron
densities outside the low-error range, there are not evident peculiarities to
produce the high values of \ff.
For NGC 6629, Mendez \etal (1992) also found \ff~ $>$1, even using different
distance and electron density.

Among galactic bulge PNe, Ha 1-31, M 2-6, M 2-30 and M 3-29
have a filling factor greater than unity.
Ha 1-31 is the only one whose electron density is not
reliable as it falls outside the low-error range.
The reason for high values
of the filling factors should reside in the observational errors.

We are aware that the available measurements of
angular diameters might not represent well the
complex morphologies of the objects considered, and
large uncertainties affect the determination
of ${\theta}$.
At a given distance, whatever the error on ${\theta}$
is, the filling factor would decrease of the same amount cubed,
since ${\varepsilon}\propto{\theta}^{-3}$. The
physical meaning is that the PN may show spatial density
fluctuations. The fact that we can overestimate (or underestimate)
the nebular radius may be balanced by the fact that
the PN is not uniformly thin or thick to the star radiation
in all directions.

Sixteen Magellanic Cloud PNe have \ff~ $>$1. Most angular diameters for
Magellanic Cloud PNe are mere upper limits, then we expected many off-range
values. In particular, we found filling factors much larger than unity when
using angular diameters from Speckle-interferometry,
which are highly unreliable (Wood 1992, private communication).
The errors or wrong determinations of $\theta$ do not propagate into errors in
\mion~, which are then very reliable for Magellanic Cloud PNe.

We estimate the general errors on ${\rm M_i}$ and ${\varepsilon}$
by looking at the errors quoted in the papers where the physical
parameters were derived:
for galactic PNe, the error estimates
on fluxes are taken from CKS92, those on electron densities
from SK89, the ones on distances from
Gathier \etal (1986) and the ones on angular diameters
from the references cited in CKS92;
the error on the galactic center distance is of the order
of $20\%$. Errors on distances
and on fluxes of extragalactic
PNe come from Kaler \& Jacoby (1990), the ones on
angular diameters from Wood \etal (1986).
Angular radii affect the calculation of ${\varepsilon}$ the most:
a $20\%$ error on the angular radius means a variation of $0.3\div0.6$
in the estimate of the filling factor. An error on angular dimensions
do not influence the ionized mass. The maximum uncertainty on ${\rm M_i}$
determined by the contributions of all observed parameters is $0.1$ \ms for
galactic PNe, and about $0.05$ \ms in the Magellanic Clouds.

\section{Conclusions and future perspectives}

The present investigation underlines the complex nature
of PNe and the various aspects that need to be taken into
account when attempting statistical studies on these objects.
Among 4 sets of galactic and extragalactic PNe at known distances,
we found that the filling factors and ionized masses spread over
wide ranges, suggesting that many different PNe types do exist.

The average
value of the filling factor is between 0.3 and 0.4 for those
nebulae whose observed parameters have errors within 20\% of their values.
We also found that
the filling factors anticorrelate with nebular dimensions
for galactic disk PNe (but less tightly than what other authors have found),
while the correlation almost disappears in galactic bulge PNe.
Since for galactic disk PNe we have used the most up-to-date
distances available, we conclude
that there is not yet evidence for a real physical correlation
among the two quantities.

The average ionized masses are between 0.1 and 0.25 \ms, depending
on the set that we consider.
The remarkable difference among ionized masses of galactic disk, galactic
bulge, and Magellanic Cloud PNe is that there is
a higher percentage of galactic disk PNe with low ionized mass.
In fact, about 90\% of disk planetaries have \mion~ $<$ 0.2 \ms, while other
sets of PNe reach this statistics only for \mion~ $<$ 0.4 \ms. In this respect,
the value \mion~=0.2\ms is an appropriate upper limit to the ionized mass
for our disk PNe, while for the other PNe this value represents
the average ionized mass.

For Magellanic Cloud PNe, where the distance is not a factor, we found tight
correlations between the filling factors and the ionized masses versus the
electron densities, as derived from forbidden line analysis.
{}From the analytical relations found, we derive that the assumptions under
Shklowsky's modified distance scale, as used in CKS92, are reliable
within the errors in the data.

Ionized masses for Magellanic Cloud PNe spread versus the core masses in the
domain predicted by the different theoretical studies; a single, sharp
relation between nebular and core mass could not be found, while the data seem
to indicate that two different trends do exist.

We plan to further expand the calculation of filling factors and ionized masses
to other samples of PNe, expecially in extragalactic environments, as more
spectroscopic data will become available, to give more statistical
significance to the results found here.

\section {Acknowledgements}

We are indebt to Alvio Renzini, Laura Greggio, James B. Kaler, and Nino Panagia
for their support, and for important comments during the development of this
research. We also thank Tim Young for the valuable help in
constructing the figures, and an anonymous referee for an important comment.

\section{References}
\par\noindent
{Acker A., Stenholm B., Tylenda R., 1989a, A\&AS 77, 487}
\par\noindent
{Acker A., K\"oppen J., Stenholm B., Jasniewicz G.,
1989b, A\&AS 80, 201}
\par\noindent
{Aller L. H., 1983, ApJ 273, 590}
\par\noindent
{Barlow M. J., 1987, MNRAS 227, 161}
\par\noindent
{Brocklehurst M., 1971, MNRAS 153, 471}
\par\noindent
{Cahn J. H., Kaler J. B., Stanghellini L., 1992,
A\&AS 94, 399 (CKS92)}
\par\noindent
{Dopita M. A., 1992, Planetary Nebulae in the Magellanic Clouds.
In: Acker A., Weinberger R. (eds.) Proc. IAU Symp. 155, Planetary
Nebulae, Kluwer Academic Publishing, Dordrecht, (in press)}
\par\noindent
{Dopita M. A., Meatheringham S. J., 1990, ApJ 356, 140}
\par\noindent
{Dopita M. A., Meatheringham S. J., Webster B. L.,
Ford H. C., 1988, ApJ 327, 639}
\par\noindent
{Gathier R., Pottasch S. R., Pel J. W., 1986, A\&A 157, 171}
\par\noindent
{Gathier R., Pottasch S. R., Goss W. M., van Gorkom J. M.,
1983, A\&A 128, 325}
\par\noindent
{Habing H. J., te Lintel Hekkert P., van der Veen W. E. C. J., 1989,
OH/IR Stars and other IRAS point sources as progenitors of
Planetary Nebulae.
In: Torres--Peimbert S. (ed.) Proc. IAU Symp. 131,
Planetary Nebulae. Kluwer Academic Publishing, Dordrecht, p. 359}
\par\noindent
{Harris W. E., 1976, AJ 81, 1095}
\par\noindent
{Iben I., Renzini A., 1983, ARA\&A 21, 271}
\par\noindent
{Jacoby G. H., Walker A. R., Ciardullo R., 1990, ApJ 365, 471}
\par\noindent
{Kaler J. B., 1970, ApJ 160, 881}
\par\noindent
{Kaler J. B., 1986, ApJ 308, 322}
\par\noindent
{Kaler J. B., Jacoby G. H., 1990, ApJ 362, 491}
\par\noindent
{Kaler J. B., Jacoby G. H., 1991, ApJ 372, 215}
\par\noindent
{Kingsburgh R. L., Barlow M. J., 1992, MNRAS 257, 317}
\par\noindent
{Kingsburgh R. L., English J., 1992, MNRAS 259, 635}
\par\noindent
{Mallik D. C. V., Peimbert M., 1988, Rev. Mex. A\&A 16, 111}
\par\noindent
{Meatheringham S. J., Dopita M. A., 1991, ApJS 75, 407}
\par\noindent
{Meatheringham S. J., Dopita M. A., Morgan D. H., 1988,
ApJ 329, 166}
\par\noindent
{Mendez R. H., Kudritzki R. P., Herrero A., 1992, A\&A 260, 329}
\par\noindent
{Mendez R. H., Kudritzki R. P., Herrero A., Husfeld D., Groth
H. G., 1988, AA 190, 113, (MEA88)}
\par\noindent
{Monk D. J., Barlow M. J., Clegg R. E. S., 1988,
MNRAS 234, 583}
\par\noindent
{O' Dell C. R., 1962, ApJ 135, 371}
\par\noindent
{Osmer P. S., 1976, ApJ 203, 352}
\par\noindent
{Osterbrock D. E., 1989, Astrophysics of Gaseous Nebulae
and Active Galactic Nuclei, University Science Book}
\par\noindent
{Osterbrock D. E., Flather E., 1959, ApJ 129, 26}
\par\noindent
{Pedreros M., 1989, AJ 98, 2146}
\par\noindent
{Peimbert M., Torres-Peimbert S., 1987, Rev. Mex. A\&A 14, 540}
\par\noindent
{Pottasch S. R., 1980, A\&A 89, 336}
\par\noindent
{Pottasch S. R., 1983. Planetary Nebulae. Reidel, Dordrecht}
\par\noindent
{Pottasch S. R., Acker A., 1989, A\&A 221, 123}
\par\noindent
{Reid M. J., 1989, The distance to the Galactic Center: $R_0$.
In: Morris M. (ed.) Proc. IAU Symp. 136,
The Center of the Galaxy. Kluwer Academic Publishing, Dordrecht, p. 37}
\par\noindent
{Renzini A., 1989, Thermal Pulses and the formation
of Planetary Nebula shells.
In: Torres--Peimbert S. (ed.) Proc. IAU Symp. 131, Planetary Nebulae.
Kluwer Academic Publishing, Dordrecht, p.391}
\par\noindent
{Renzini A., Voli M., 1981, A\&A 94, 175, (RV81)}
\par\noindent
{Rich R. M., 1991, The Stellar Population of the Galactic Bulge.
In: Barbuy B., Renzini A. (eds.) Proc. IAU Symp. 149,
Stellar Populations of Galaxies, Kluwer Academic Publishing, Dordrecht,
p. 29}
\par\noindent
{Sanduleak N., Mac Connell D. J., Philip A. G. D.,
1978, PASP 90, 621, (SMP78)}
\par\noindent
{Schmidt--Voigt M., K\"oppen J., 1987a, A\&A 174, 211}
\par\noindent
{Schmidt--Voigt M., K\"oppen J., 1987b, A\&A 174, 223}
\par\noindent
{Seaton M. J., 1966, MNRAS 132, 113}
\par\noindent
{Shaw R. A., Kaler J. B., 1982, ApJS 69, 495}
\par\noindent
{Stanghellini L., Kaler J. B., 1989, ApJ 343, 811, (SK89)}
\par\noindent
{Stanghellini L., Corradi R. L. M., Schwarz H. E., 1993, A\&A (in press)}
\par\noindent
{Stasi\'nska G., Tylenda R., Acker A., Stenholm B.,
1991, A\&A 247, 173, (STAS91)}
\par\noindent
{Torres-Peimbert S., Peimbert M., 1977, Rev. Mex. A\&A 2, 181}
\par\noindent
{Tylenda R., Acker A., Gleizes F., Stenholm B., 1989, A\&AS 77, 39}
\par\noindent
{Webster B. L., 1969, MNRAS 143, 79}
\par\noindent
{Webster B. L., 1976, MNRAS 174, 513}
\par\noindent
{Webster B. L., 1983, PASP 95, 610}
\par\noindent
{Vassiliadis E., Wood P. R., 1992, ApJ, in press (VW92)}
\par\noindent
{Wood P. R., Bessell M. S., Dopita M. A., 1986, ApJ 311, 632}
\par\noindent
{Wood P. R., Meatheringham S. J., Dopita M. A., Morgan D. H., 1987,
ApJ 320, 178, (WMDM)}
\end{document}